\documentclass[%
 preprint, 
 amsmath,amssymb,
 aps, physrev,
]{revtex4-2}

\usepackage{graphicx}% Include figure files
\usepackage{dcolumn}% Align table columns on decimal point
\usepackage{bm}% bold math
\usepackage{color}
\usepackage{afterpage}
\usepackage{caption}
\usepackage{physics}
\usepackage{comment}
\usepackage{float}

\begin{document}

%\preprint{APS/123-QED}

\title{Stability Control of Metastable States as a Unified Mechanism for Flexible Temporal Modulation in Cognitive Processing}

\author{Tomoki Kurikawa}
 \email{kurikawa@fun.ac.jp}
 \affiliation{Department of Complex and Intelligent Systems, Future University Hakodate, 116-2 Kamedanakano-cho,
Hakodate, Hokkaido, 041-8655, Japan}%Lines break automatically or can be forced with \\

\author{Kunihiko Kaneko}%
\affiliation{
 The Niels Bohr Institute, University of Copenhagen, Blegdamsvej 17, Copenhagen, 2100-DK, Denmark
 }
\affiliation{
 Center for Complex Systems Biology, Universal Biology Institute, University of Tokyo, Komaba, Tokyo 153-8902, Japan
 }
\date{\today}% It is always \today, today,

\begin{abstract}
Flexible modulation of temporal dynamics in neural sequences underlies many cognitive processes. For instance, we can adaptively change the speed of motor sequences and speech. While such flexibility is influenced by various  factors such as attention and context, the common neural mechanisms responsible for this modulation remain poorly understood.
We developed a biologically plausible neural network model that incorporates neurons with multiple timescales and Hebbian learning rules. This model is capable of generating simple sequential patterns as well as performing delayed match-to-sample (DMS) tasks that require the retention of stimulus identity. Fast neural dynamics establish metastable states, while slow neural dynamics maintain task-relevant information and modulate the stability of these states to enable temporal processing.
We systematically analyzed how factors such as neuronal gain, external input strength (contextual cues), and task difficulty influence the temporal properties of neural activity sequences—specifically, dwell time within patterns and transition times between successive patterns. We found that these factors flexibly modulate the stability of metastable states.
Our findings provide a unified mechanism for understanding various forms of temporal modulation and suggest a novel computational role for neural timescale diversity in dynamically adapting cognitive performance to changing environmental demands.

\end{abstract} %%%%%%%%%

\maketitle
\section*{Author Summary}
The brain often uses sequences of neural activity to perform complex cognitive tasks such as recognizing speech, making decisions, or holding information in working memory. These sequences can speed up or slow down depending on factors like attention, task difficulty, or expectations—but how the brain controls this timing remains unclear. 
In this study, we built a biologically plausible model of a neural network that includes both fast and slow neurons and learns tasks through simple, realistic rules. 
We show that the slow neurons can hold onto past information and control how long the network activity stays in each state of a neural sequence. 
This control depends on the stability of each state, which is influenced by factors such as the external input strength, task difficulty, and top-down modulation. 
Our model coherently explains a variety of experimental findings and provides a unified theory for how the brain might flexibly adjust the speed of thought by taking advantage of diverse timescales of neural activity.

\section*{Introduction}
Cognitive functions such as speech recognition, working memory\cite{Ponce-Alvarez2012,Stokes2013,Taghia2018}, perceptions\cite{Jones2007,Miller2010} and context-dependent decision-making\cite{Bollimunta2012,Wimmer2020,Benozzo2021} rely on the brain's ability to flexibly adjust the timing of sequential neural activity.
These temporal sequences are not rigidly fixed; instead, they adapt their pace in response to contextual factors such as arousal\cite{Engel2016,Wyrick2021}, task difficulty\cite{Benozzo2021,Ponce-Alvarez2012}, and expectation\cite{Mazzucato2019,Wyrick2021}.
For example, the duration of metastable neural states tends to be longer in highly aroused states\cite{Engel2016}, shorter in high-expectation condition\cite{Mazzucato2019,Wyrick2021}, and more prolonged during incorrect responses in perceptual discrimination tasks\cite{Benozzo2021}. 
Likewise, transitions between patterns are delayed in difficult trials\cite{Ponce-Alvarez2012}, and expedited when stimuli are anticipated\cite{Mazzucato2019,Wyrick2021}. 
These findings suggest that flexible temporal modulation is a general feature of neural computation, yet the underlying mechanisms remain elusive.

Various models have been proposed to explain the temporal modulation of sequential patterns in neural networks.
In these models, the influence of several factors is investigated, such as external input and the gain in neuronal activation functions \cite{Gillett2024,Murray2017,Wyrick2021,Mazzucato2019}.
For example, the Hopfield network with the connections formed by the correlation between the successive patterns\cite{Gillett2024} and the cluster network model with the anti-Hebbian learning rule\cite{Murray2017} demonstrated that change in external input strength can modulate the speed of pattern sequence generation.
Similarly, the gain parameter was shown to alter the speed of sequential patterns in the cluster network model\cite{Wyrick2021,Mazzucato2019}.

While these models contribute to our understanding of simple sequential patterns, there are significant limitations.
Specifically, the subsequent transition pattern is completely or stochastically determined by the preceding pattern, meaning that these sequential patterns are neither input modulated nor history-dependent.
Consequently, these models fail to capture mechanisms underlying information processing in most cognitive tasks, such as sensory-motor coupling tasks and working memory tasks, which require temporal patterns to be input-dependent and/or history-dependent.
Furthermore, these models typically focus on single factors regulating temporal modulations.

Besides these biologically plausible models, some studies employed advanced machine learning techniques on recurrent neural networks to investigate neural dynamics in cognitive tasks\cite{Hardy2018,Wang2018,Zhou2023} and demonstrated that input strength modulated the speed of neural state transitions. 
However, these models come up against significant difficulties: they require training under multiple speed conditions to achieve continuous speed modulation, and their learning mechanisms lack a clear biological basis.
Consequently, the underlying biological principles of speed control remain poorly understood.

To overcome these limitations and explore a mechanism underlying the flexible temporal modulation, this study develops a neural network model employing a simple Hebbian rule\cite{Kurikawa2013,Kurikawa2020} and neurons with multiple timescales\cite{Kurikawa2020a,Kurikawa2021b,Maes2021,Perez-Nieves2021,Kurikawa2024}, which is widely observed in the neural systems\cite{Bernacchia2011,Wasmuht2018,Scott2017,Cavanagh2018,Cavanagh2016}.
This model generates input-dependent and history-dependent sequences and successfully performs working memory tasks.
Using this model, we explore how several factors, such as the strength of the context cue\cite{Wang2018,Hardy2018}, the gain of the activation function of neurons, and the task difficulty in a working memory task, change the temporal sequential patterns. 
Our findings demonstrate that a wide range of factors—including arousal, task difficulty, and external input strength—can modulate the stability of metastable states in neural sequences, thereby controlling their speed.
We propose a unified mechanism of flexible temporal modulation based on multiple neural timescale: it is achieved through changes in the transition time and the dwell time on the pattern, mediated by stability modulation of metastable states.
Further, these results are consistent with experimental observations.

%attractor dynamicsの実験的な説明いる？

%個別には色々知られている。
%Sequenceのでき方。Jazayeri, Buonomano…

\section*{Model and Methods}

\subsection*{A neural network model}
We consider two types of temporal processing in the same neural network model: simple sequence generation and history-dependent processing such as delayed match-to-sample (DMS) tasks.

To achieve this, we use the two-population model with different timescales (Fig. \ref{fig:image}A), one with $N$ fast neurons and the other with $N$ slow neurons, denoted as $X$ and $Y$, respectively, as in the previous studies\cite{Kurikawa2020a}.
$X$ receives an external input $\bm{\eta}$, and $Y$ receives the output of $X$ and provides input to $X$ recursively.
The neural activities $x_i$ in $X$ and $y_i$ in $Y$ evolve according to the following equation:
\begin{align}
  \tau_{x}\dot{x_{i}} &=& \tanh{(\beta I_{i})} - x_{i},  \label{eq:neuro-dyn}\\ 
  \tau_{y}\dot{y_{i}} &=& \tanh(\beta_y x_{i}) - y_{i},  \label{eq:neuro-dyn-slow}\\ 
  I_i  &=& u_i  +  \gamma_{y} r_i +\gamma \eta_{i},  \label{eq:input}
\end{align}
where $u_{i} = \sum_{j \neq i}^{N} J_{ij}^{X} x_{j}$, $r_{i} = \sum_{j}^{N} J_{ij}^{XY} y_{j}$.
$J_{ij}^{X}$ is a recurrent connection from the $j$-th to the $i$-th neuron in $X$, and $J^{XY}_{ij}$ is a connection from the $j$-th neuron in $Y$ to the $i$-th neuron in $X$.
 $J^{X}$ is a fully connected connectivity with no self-connections, and its mean and variance are set to zero and $1/N$, respectively, while $J^{XY}$ is a sparse random connectivity with path density $0.1$, and the mean and variance of non-zero connections are set to zero and $49/N$, respectively.
$\gamma$ and $\gamma_y$ are the strength of the external input and that of input to $X$ from $Y$,  respectively.
$\tau$ and $\tau_y$ are the timescales of the fast and slow neurons.
We set $N=100, \beta=2, \beta_y=20, \tau_x=1, \gamma=1, \gamma_y=0.5$ and $\tau_y=100$.

\subsection*{Simple sequence generation task}

The activity pattern of $X$ is the output of this model.
The model is needed to match its output to target sequential patterns $\boldsymbol{\xi}^{\mu}$ $(\mu=1,\cdots, M)$ in the presence of $\boldsymbol{\eta}$,
i.e. pseudo-attractors in $X$ that match $\boldsymbol{\xi}^{\mu}$ should be formed through learning.
The $i$-th element of a targeted pattern  $\xi^{\mu}_i$ is assigned to the $i$-th neuron in $X$, and its value is randomly sampled according to the probability $P[ \xi^{\mu}_{i}=\pm 1]=1/2$.
The task signal $\bm{\eta}^\mu_i$ is injected into the $i$-th neuron in $X$, randomly sampled according to $P[\eta_i=\pm 1]=1/2$. 
The number of patterns in sequence, $M$, is set to $5$ in this study.

Only $J^{X}$ changes to match the output to the target (one of the sequential patterns) according to the following equation: 
\begin{align}
	\tau_{syn}\dot{J_{ij}^{X}} &=& (1/N)(\xi_i - x_i)(x_j - u_i J_{ij}^X), \label{eq:lrn_dyn}
\end{align}
where $\tau_{syn}$ is the learning speed (set at $100$).
This learning rule comprises a combination of a Hebbian term between the target and the activity of the presynaptic neuron, and an anti-Hebbian term between the pre- and post-synaptic neurons with a decay term $u_i J_{ij}^X$ for normalization.
This form satisfies locality in connectivity and is biologically plausible \cite{Kurikawa2020a}.
We previously applied this learning rule to a single network of $X$, and demonstrated that the network learns static input/output maps (not sequences) i.e., $M=1$ \cite{Kurikawa2013,Kurikawa2016,Kurikawa2020}.
However, in the single network model, generating a sequence ($M \geq 2$) was not possible.
In the network model with two sub-networks, as in the present study, there are two inputs to $X$, 
one from the external task signal $\boldsymbol{\eta}$, and the other from $Y$, which stores previous information.
Thus, the network can generate a pattern depending not only on the current input (task) signal, but also on the previous patterns, which is necessary to perform a delayed match-to-sample task, as described below.

A learning step of a single pattern is completed when the neural dynamics meet the following two criteria:
$\boldsymbol{x}$ is sufficiently close to the target pattern,
i.e., $m_{\mu}^x \equiv \Sigma_i x_i (\xi^1_{\mu})_i/N > 0.9$,
and also $\boldsymbol{y}$ is sufficiently close to $\boldsymbol{x}$, 
i.e., $\Sigma_i x_i y_i /N > 0.5$.
After the completion of one learning step, a new pattern $\boldsymbol{\xi}^2$ is assigned instead of $\boldsymbol{\xi}^1$ with a perturbation of fast variables $x_i$, by multiplying a random number uniformly sampled between zero to one.
We execute these steps sequentially from $\mu=1$ to $M$ to learn a sequence.
Each time, after learning the sequence once, we test whether the network generates the sequence; 
we run the neural dynamics without changing the connections and check if the sequence is successfully recalled four times in a row.
When the sequence is generated correctly, the learning and the neural dynamics are completed.

\subsection*{Delayed match to sample (DMS) tasks}
For training DMS tasks (as illustrated in Fig. \ref{fig:DMS_image}A), the learning rule is the same as that for the sequence generations, except for the duration in which the connectivity is modulated and input/target patterns applied to the network.
The parameters are also the same as those used in the sequence generations, except  $\beta=1.5$ and $\tau_{syn}=250$.

The input patterns are implemented as follows.
We have two input patterns, A and B, that are generated in the same manner as in the simple sequence generations.
In this task, either input pattern (chosen at random) is applied to the fast variables for 30 time units, followed by a delay of 30 time units.
Then either input pattern (again chosen at random) is applied.

The network is trained to generate "Match" or "Non-Match" pattern through the learning process, depending on whether the two successive inputs are the same.  
For this purpose, we set two target patterns $\bm{\xi}^{Match, Non-Match}$, corresponding to "Match" and "Non-Match" pattern,  that are composed of $N_\text{trgt}=50$ neurons.
A half of the fast neurons ($i=1,2,\ldots 50$) are assigned to the output of the network to be compared with the target pattern.
These two target patterns are generated in the same manner as in the sequence generations.
In the present model, the learning phase (given by Eq. \ref{eq:lrn_dyn}) requires a target signal for each neuron and we have dummy patterns composed of $N-N_\text{trgt}$ fast neurons ($i=51,52,\ldots 100$) that are generated randomly.
In this task, the model should output the Match pattern, namely the targeted activity pattern of $50$ fast neurons ($i=1,2,\ldots 50$),  when the two successive input patterns are the same, otherwise it should output the Non-Match pattern.
To visualize the response of the network, we formally implement "Match" and "Non-Match" neurons whose activities are determined by $x_\text{Match}=\Sigma_i^{N_\text{trgt}} x_i \xi^{Match}_i/N_\text{trgt}$ and $x_\text{Non-Match}=\Sigma_i^{N_\text{trgt}} x_i \xi^{Non-Match}_i/N_\text{trgt}$, respectively.

We run the learning process simultaneously with the neural dynamics only after the delay time (the onset of the second stimulus), otherwise we do not run it.
Due to the change in connectivity, the activities of the fast variables learn to converge to the correct response Match or Non-Match.
In this model, the slow variables retain the information about the first applied stimulus even after the delay time, allowing the network to generate the correct response through such a simple learning rule.

Finally, we describe the conditions for completing learning in one trial.
A decision period is set at two time units after the delay period.
When either $x_\text{Match}$ or $x_\text{Non-Match}$ exceeds $0.9$ during the decision period, the learning and neural dynamics are completed and the next trial starts. 
Here, these two additional time units are set to prevent $x_\text{Match}$ or $x_\text{Non-Match}$ from remaining high for a long time before the onset of the second stimulus.

\subsection*{Measuring the transition and dwell time}

First, we identify the time $t_{in}$ at which the pattern of fast neurons first goes to a pattern $A$ beyond the threshold $\theta$, defined as $t_{in} \stackrel{\mathrm{def}}{=} \underset{t} {\operatorname{argmin}} \Sigma_{i\leq N} \xi_i x_i(t)/N > \theta$ for a simple sequence generation and $t_{in} \stackrel{\mathrm{def}}{=} \underset{t} {\operatorname{argmin}}\Sigma_{i\leq N_{trgt}} \xi_i x_i(t)/N_{trgt} > \theta$ for the DMS task.
Also, the time $t_{out}$ at which the pattern departs from the pattern $A$ is defined as $t_{out}  \stackrel{\mathrm{def}}{=} \underset{t_{in} \leq t }{\operatorname{argmin}} \Sigma_{i\leq N} \xi_i x_i(t)/N < \theta$.
In this paper, $\theta=0.8$ is used.

Then, for the simple sequence $A \rightarrow B$, the dwell time of the pattern $A$ is defined as $t_{out} - t_{in}$, as shown in Fig. \ref{fig:gain}D.
Similary, the transient time to $B$ is defined as difference between $t_{in}$ for $B$ and $t_{out}$ for $A$.
For DMS task, where there are no sequential patterns, we define the transition time (or so-called reaction time) to the final decision pattern, Macth or Non-Match pattern,  as the difference between $t_{in}$ for the final decision pattern and the onset time of the second stimulus.

\subsection*{Measurement of the stability}

Here, we introduce the equation given in Eq. \eqref{eq:deltam}.
Now, for the sake of simplicity, $\tanh(\beta x)$ is denoted as $\phi(x)$.
By taking the second order of Taylor expansion,
we obtain
\begin{align}
    \Delta m 
    &\sim (\boldsymbol{\xi^{\mu}} )^T (\phi( I_{\mu}+\boldsymbol{\epsilon})-\phi(I_{\mu}))/N \\
    &\sim  (\boldsymbol{\xi^{\mu}} )^T (\phi'(I_{\mu}) \boldsymbol{\epsilon} + \frac{1}{2}\phi''(I_{\mu}) \boldsymbol{\epsilon}^2)/N.
\end{align}
The first linear term is averaged out, because the mean value of $\boldsymbol{\epsilon}_i$ is zero and $\phi'(I_{\mu})_i$ is uncorrelated with $\boldsymbol{\epsilon}_i$.
Given that the second derivative $\phi^{''}(x)$ equals $-2 \beta^2 (1-\phi(x)^2)\phi (x)$,
the second term is represented by 
\begin{align}
    \text{second term} &= -\beta^2 \Sigma_i \xi_i^{\mu}   (1-\phi(I_{\mu})_i^2)\phi( I_{\mu})_i \epsilon_i^2 /N  \\
    &\sim -\beta^2 \{\Sigma_i \xi_i^{\mu}   (1-\phi(I_{\mu})_i^2)\phi(I_{\mu})_i /N \} \{ \Sigma_i  \epsilon_i^2 /N\}.
\end{align}
Here, $\phi(I_{\mu}) $ can be represented by $\xi_i ( 1- k_i)$  with a small value $0<k_i \ll 1$.
By using this representation, we transform $ (1-\phi_i^2)\phi_i = 2\xi_i^{\mu} k_i + O(k_i^2) \sim 2 \xi_i^{\mu} (1- \xi_i^{\mu}\phi_i)$ and obtain
\begin{align}
    \Sigma_i \xi_i^{\mu}   (1-\phi(I_{\mu})_i^2)\phi(I_{\mu})_i /N  &= \Sigma_i 2 (\xi_i^{\mu})^2 (1-\xi_i^{\mu}\phi(I_{\mu})_i)/N \\
    &= 2(1-s_A).
\end{align}
Thus, the second term is represented by
\begin{align}
    \text{second term} &= -2\beta^2 (1-s_A) \sigma^2.
\end{align}
In total, $\Delta m = -2\beta^2 (1-s_A) \sigma^2$.

\newpage
\begin{figure}[h]
    \centering
    \includegraphics[width=0.95\linewidth]{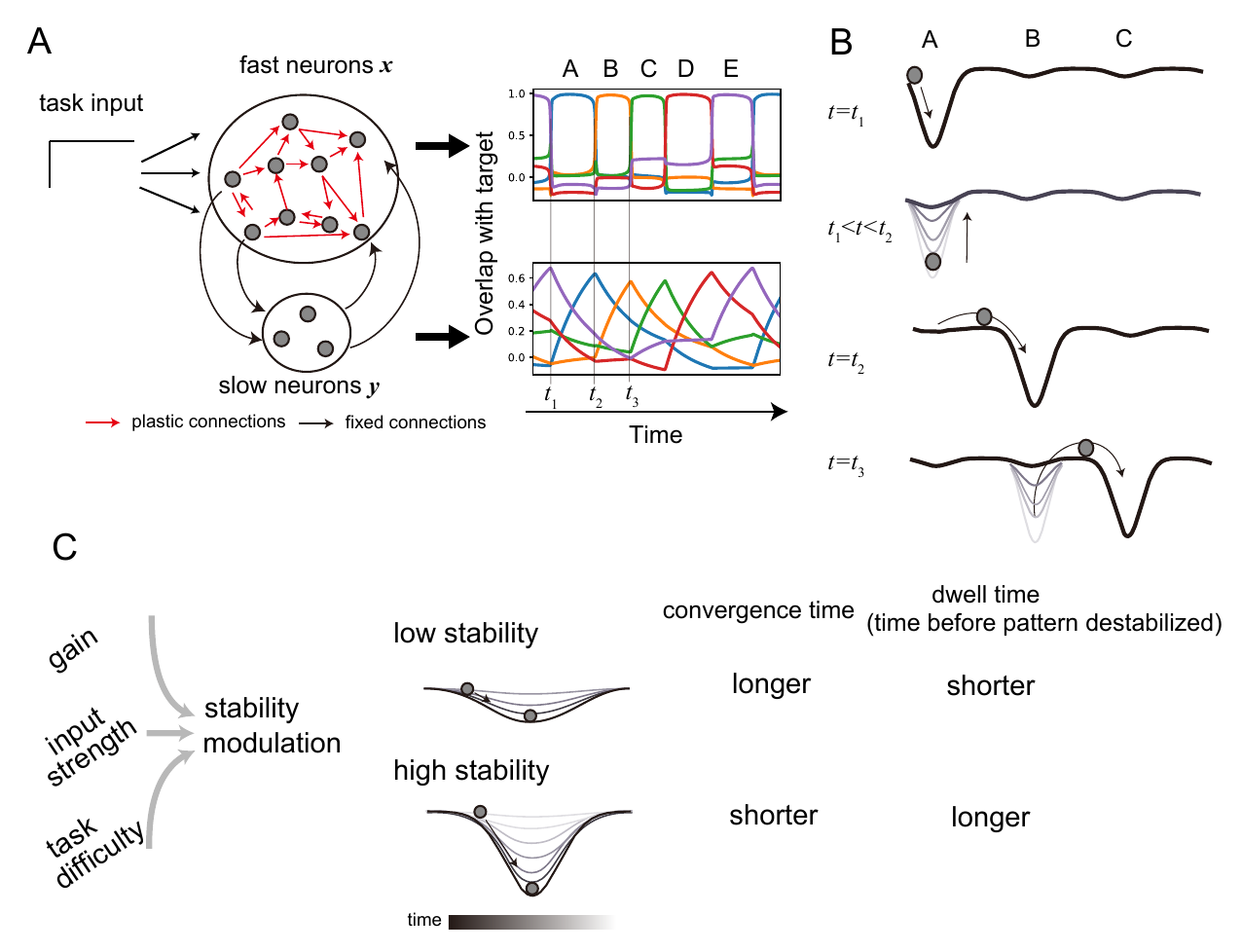}
    \caption{Schematic image of our study. 
    A: Two-population networks generating sequential patterns. Left: schematic image of our model. Right: the sequence generation with overlaps of the neural activties with targets. Different colors show the overlaps with different targets.
    B: Schematic landscape of transitions from one to another patterns in sequence.
    After the transition of the state of the fast variables to $A$ at $t=t_1$, the stability becomes unstable through the change in the slow variables $t_1<t<t_2$.
    The neural state transits from A to B when the stability of A is broken.
    C: Several factors control the neural state speed through modulation of stability of sequential patterns.
    For low stability of a pattern, the convergence time to the pattern is long (the convergence speed is smaller) and the dwell time (the time before the pattern becomes unstable) is shorter, 
    and the opposite for high stability. }
    \label{fig:image}
\end{figure}

\newpage
\section*{Results}

\subsection*{Generation of simple sequential patterns in the fast variables}
We analyze the common mechanism for flexible speed control in several temporal information processing: simple sequence generation and DMS task.
Here, we introduce our proposed mechanism with simple sequence generation as an example. 
After learning the target sequence, the model outputs it.
The output of the model is represented by the activity pattern of the fast neurons $X$, while the target pattern is the same dimensional vector as the output (see A neural network model in 
Methods for details).

We show how the sequence patterns are generated after learning (the mechanism for the generation of sequences was analyzed thoroughly in the previous studies\cite{Kurikawa2020a,Kurikawa2021b}.
An example of a recall process after learning of the target sequence (patterns A, B, C, D, and E) is plotted with the overlap of the neural activities with the patterns in Fig. \ref{fig:image}A.
Here, the overlaps are defined as $\Sigma_i \xi_i x_i/N$, where $\xi_i$ and $x_i$ are the $i$-th element of the target pattern ($A,B,\ldots, E$) and the $i$-th fast neuron activity.
After learning, the connections are fixed in the recall of sequential patterns.
The initial states of the fast variables are set at random values sampled from a uniform distribution of -1 to 1, 
whereas the slow variables are set to the final state state of the pattern E to start the sequence at the pattern A.
The model outputs correctly the target sequence patterns A, B, C, D, and E iteratively.
Note that the transitions between patterns occur spontaneously without any external operation
through interplay between the fast and slow neural dynamics.

In our model, the transition between the patterns in the fast variables is caused by the slow variables.
When focusing on the slow variables dynamics in Fig. \ref{fig:image}A, their overlaps with the target patterns follow those of the fast dynamics;
The overlap of the fast variables with the pattern A surges to almost unity at $t=t_1$ and then that of the slow variables with the same pattern slowly increases according to their dynamics Eq. \ref{eq:neuro-dyn-slow}, in which the slow dynamics mirror the fast ones.
When the overlap of the slow variables reaches $0.6$ at $t=t_2$, the transition from the pattern A to B in the fast variables occurs.
Then, the overlap of the slow variables with the pattern B increases.
Pattern A remains stable throughout the time span as the slow variables gradually approach it, whereas it becomes unstable when the slow variables move very close to A.
Namely, the slow variables work as a bifurcation parameter to stabilize or destabilize the pattern of the fast variables, as illustrated in Fig. \ref{fig:image}B, although the slow variables themselves are driven by the fast variables (see the detailed analysis of the bifurcation diagram in \cite{Kurikawa2020}) .

This model shows a tentative mechanism in which the stability of a pattern in the sequence could regulate the dwell time of patterns and the convergence time to a pattern.
The dwell time of a pattern is the time during which the neural state of the fast variables stays on the pattern,
while the convergence to a pattern is given by the time required for the fast neural state to converge to the pattern.
If the stability of a pattern is enhanced, as shown in Fig. \ref{fig:image}C, the range of the parameter (here, the slow variables) in which the pattern is stable is increased.
Thus, the pattern remains stable even when the slow overlap exceeds the value at which the pattern becomes unstable for the normal stability. 
This prolongs the time before the pattern becomes unstable for the high stability, meaning the longer dwell time of the pattern.
In contrast, the higher stability leads to the shorter convergence time, because the convergence speed to one pattern is faster.
Oppositely, if the stability is weakened, the dwell time would be shorter and the convergence time longer.

We hypothesize that diverse factors, such as the gain factor, top-down input, and difficulty of the DMS task (see Fig. \ref{fig:DMS_image}), vary the stability of the patterns in the sequence, realizing the flexible speed modulation of the sequence and the change in transition time in the DMS task via change in the dwell time and the convergence time.
In the following part, we analyze the neural dynamics in our model to uncover a possible relation between the dwell and transition times and the stability of patterns in the sequence and the DMS task.

%%%%%%%%%%%%%%%%%%%%%%%%%%%%%%%%%%%%%%%%%%%%%%%%%%%%%%%%%%%%%%%%%%%%%%%%%%%%%%%%%%%%%%%%%%%
%%%%%%%%%%%%%%%%%%%%%%%%%%%%%%%%%%%%%%%%%%%%%%%%%%%%%%%%%%%%%%%%%%%%%%%%%%%%%%%%%%%%%%%%%%%

\subsection*{Gain factor modulates the generating speed of sequential patterns}

Now, we explore how the gain parameter modulates the speed of the sequence.
Actually, the gain parameter is considered to be related to neural modulators\cite{Shine2021,Cools2022,Mather2016}, such as ACh and Noradrenaline, that are associated with the cognitive states like attention\cite{Thiele2018} and arousal state\cite{Teles-GriloRuivo2017,Lee2018}. 
The gain parameter in the recall process after learning is distinguished from that used in the learning process: The network was trained to generate the sequence with $\beta=3$, denoted as $\beta_\text{lrn}$, and then, in the  recall, we analyze the neural dynamics by changing the gain parameter.

\begin{figure*}[h]
    %\centering
    \includegraphics[width=0.85\linewidth]{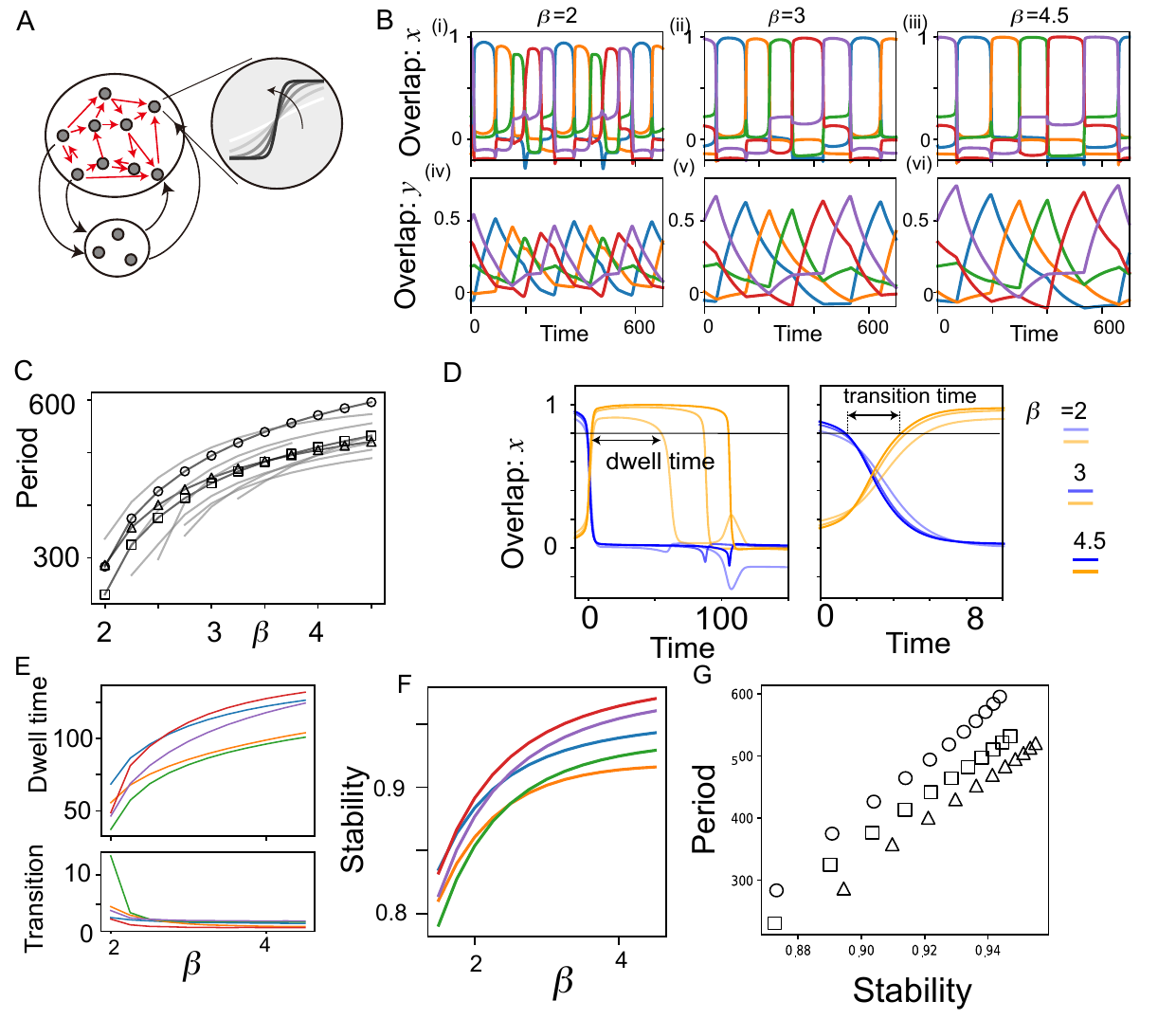}
    \caption{Change in the gain parameter $\beta$ modulates the neural dynamics.
    A: Schematic image of the change in the gain parameter in fast neurons.
    B: The neural dynamics for different values of $\beta$ are plotted with their overlaps with the patterns, A, B, C, D, and E.
    The upper panels exhibit the overlaps of the fast variables, while the bottom panels exhibit those of the slow.
    Different colors show the overlaps with different patterns, as shown in Fig. \ref{fig:image}A.
    The color code is the same across D, E, and F.
    C: The period of the one sequence is plotted by varying $\beta$. Ten different lines show different realizations of the networks learning different sequences.
    Circles, triangles, and squares represent the three realizations that are the same as shown in G.
    D: Examples of the dwell time (left panel) and the transition time (right panel). 
    The threshold value is set at $0.8$.
    The right panel exhibits the expansion of the left.
    E: The dwell time for each pattern of the network shown in B is plotted for varying $\beta$ in the top panel, as well as the transition time in the bottom panel. 
    F: The stability for each pattern  is plotted in the same way as in E.
    G: The relation between the stability and period for the three networks is plotted. 
    The markers correspond to the same marker shown in C.
    }
    \label{fig:gain}
\end{figure*}

The gain parameter $\beta$ changes the period of the sequence.
Figure \ref{fig:gain}B illustrates the representative neural dynamics used by their overlap with five learned patterns, demonstrating the neural dynamics with varying gain $\beta$.
For $\beta=\beta_\text{lrn}=3$, the five patterns in the fast variables are generated sequentially, 
where the period of the five-pattern generation is about $500$ unit-time.
For $\beta=2$, the period is shorter than that for $\beta=3$, whereas for $\beta=4.5$, the period is longer.
For a much smaller value of $\beta$ ($\beta < 2$), the network fails to generate the sequence.
When $\beta$  increases (for $\beta \geq 2$),  the period monotonically increases as shown in Fig. \ref{fig:gain}C.
We also confirm such $\beta$ dependence is observed in other realizations of the networks that have learned different patterns in the sequences.

\clearpage

The time when the pattern in the sequence is unstable depends on $\beta$.
For $\beta=3$, the value of the overlap of the slow variables with the pattern A reaches 0.6 around $t=150$, and simultaneously the fast neural state switches the pattern A to B, as shown in Fig. \ref{fig:gain}B(v).
For the smaller value of $\beta(=2)$, the value of the overlap at the transition point is lower,
while it is higher for the larger value of $\beta(=4.5)$, suggesting that pattern A is more stable for the larger value of $\beta$, and consequently the slow variables approach pattern A more closely, resulting in a longer time until A becomes unstable.

To test if the speed is controlled by the stability change, we investigate the relation between the stability of patterns and the dwell time, as well as the transition time.
First, we measured the dwell time on each pattern.
As shown in Fig. \ref{fig:gain}D, the dwell time of a pattern is defined as the interval between the time when the overlap of the fast variables with the pattern exceeds a threshold (here, $0.8$) and the time when the overlap falls below the threshold (see Measuring transition and dwell time in Methods).
The transition time from the pattern A to B is defined as the interval between the time when the overlap with the pattern A falls below the threshold and that when the overlap with the pattern B exceeds it.
Figure \ref{fig:gain} E exhibits the dwell time and transition time against the value of $\beta$.
The dwell times in all patterns monotonically increase, while the transition time decreases.
The transition time is much smaller than the dwell time, so that the period of the sequence is dominantly determined by the dwell time.

Next, we measured the stability of the pattern.
Here, we defined and calculated the stability factor $s_A (\beta)$ of the pattern A (corresponding to $\bm{\eta}^{\mu}$) in the sequence by
\begin{equation}
     s_A = (\bm{\xi}^\mu)^T \tanh(\beta I^\mu),
\end{equation}
where $I^{\mu}=J\bm{\xi}^{\mu}  +  \gamma_{y} J_{XY} \bm{y_0}^{\mu} +\gamma \bm{\eta}$ is the input current to the fast variable as in eq. \ref{eq:neuro-dyn} with replacing $x(t)$ and $y(t)$ by $\bm{\xi}^{\mu}$ and $\bm{y_0}^{\mu}$, respectively.
$\tanh(\beta I^\mu)$ represents the $n$ dimensional vector, $i$-th component of which is $\tanh(\beta I^\mu_i)$.
$\bm{y_0}^{\mu}$ is the pattern of the slow variables at the time when the overlap of the fast variables with $\bm{\xi}^{\mu}$ takes a peak value for $\beta=\beta_\text{lrn}=3$.
Thus, the stability is calculated when the slow variables are far from the stability-break point of $\bm{\xi}^\mu$.
Note that $-1 \leq s_A \leq 1$.
The stability $s_A$ quantifies the resilience to a perturbed input as follows:
$s_A$ represents approximately the fixed point of the dynamics  Eq. \ref{eq:neuro-dyn}
 projected to the target when $\boldsymbol{y}$ is clamped to $\boldsymbol{y_0}^{\mu}$.
 In fact, with $m_\mu=(\bm{\xi}^\mu)^T \bm{x}/N$, the fixed point denoted by $m_{\mu}^\text{FP}$ is obtained as 
 \begin{align}
     m_{\mu}^{FP} = (\boldsymbol{\xi^{\mu}} )^T \text{tanh}(J\bm{x}^{FP}  +  \gamma_{y} J_{XY} \bm{y_0}^{\mu} +\gamma \bm{\eta})/N,
 \end{align}
 where $\boldsymbol{x}^{\text{FP}}$ is the neural state of the fixed point, namely $m_{\mu}^\text{FP}=(\bm{\xi}^\mu)^T \bm{x}^\text{FP}/N$.
 By assuming that $\boldsymbol{x}^{\text{FP}} \sim \boldsymbol{\xi^{\mu}}$ after learning, $m_{\mu}^{\text{FP}} \sim s_A$ follows.
Now, we apply a perturbed input $\boldsymbol{\epsilon}$  as $\gamma \boldsymbol{\eta} \rightarrow \gamma \boldsymbol{\eta}  + \boldsymbol{\epsilon}$, and then the fixed point is modified due to the perturbation, where the mean and variance of the perturbation across elements are zero and $\sigma^2$.
The modification of the fixed point, termed $\Delta m$, satisfies that
\begin{align}
    \Delta m 
    &=-2\beta^2 (1- s_A) \sigma^2. \label{eq:deltam} 
\end{align}
See Methods for details.
Thus,  for the higher value of $s_A$, the modulation of the fixed point due to the perturbation is small.  
$s_A$ is calculated as a function of $\beta$ once $y_0$ is measured for $\beta=3$, without further measuring actual neural dynamics for any value of $\beta$.

Figure \ref{fig:gain}F shows the stability of the five patterns in the sequence as a function of $\beta$.
We find that the stability shows a similar behavior as the dwell time and the period of the sequence.
Actually, the stability averaged over patterns is highly correlated with the period as shown in Fig. \ref{fig:gain}G.
Such a high correlation between the stability and the period is observed in other realizations of the networks after learning different patterns.
It indicates that we can predict the period for any value of $\beta$ by calculating the stability without observing the actual neural dynamics per se.
These results show that the gain parameter modulates the stability of the patterns, leading to change in the period, as consistent with our hypothesis.

%%%%%%%%%%%%%%%%%%%%%%%%%%%%%%%%%%%%%%%%%%%%%%%%%%%%%%%%%%%%%%%%%%%%%%%%%%%%%%%%%%%%%%%%%%%
%%%%%%%%%%%%%%%%%%%%%%%%%%%%%%%%%%%%%%%%%%%%%%%%%%%%%%%%%%%%%%%%%%%%%%%%%%%%%%%%%%%%%%%%%%%
\begin{figure*}[h]
    %\centering
    \includegraphics[width=0.9\linewidth]{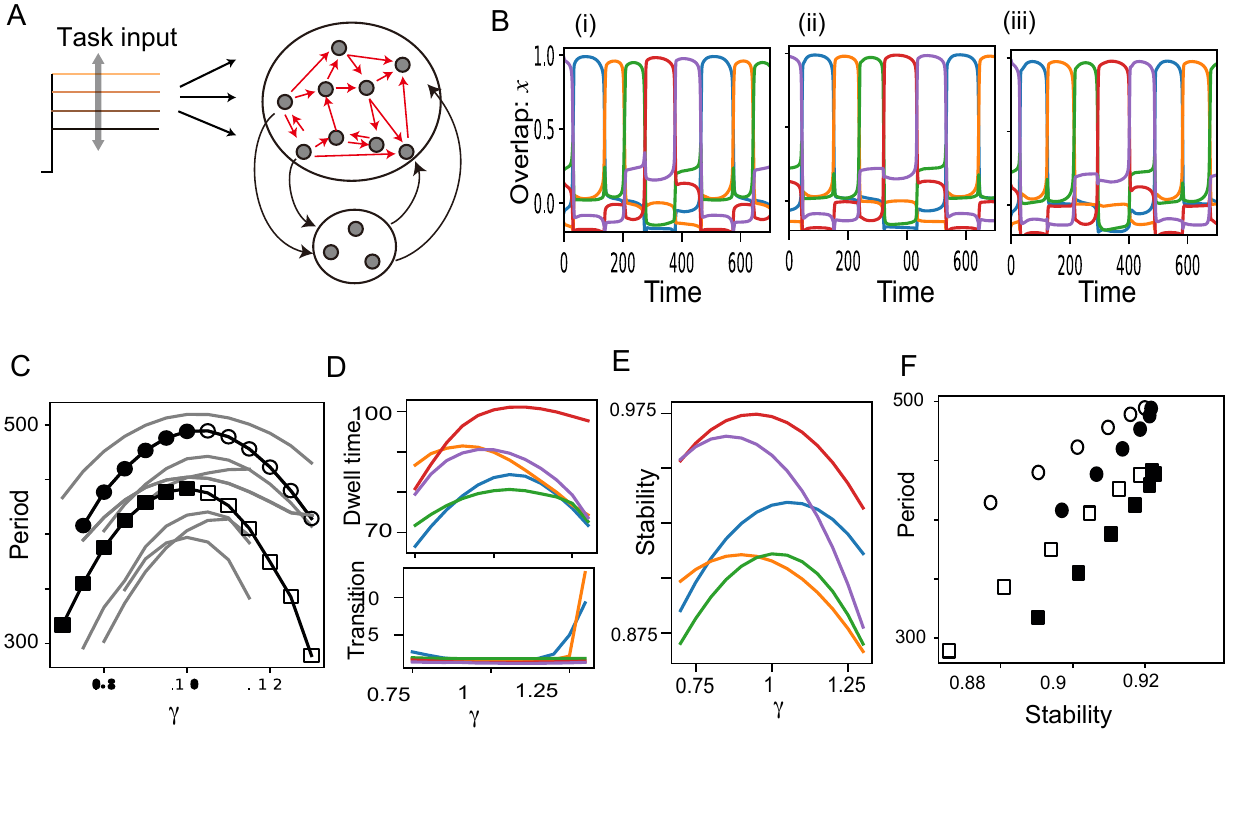}
    \caption{Change in task input strength $\gamma$ shown in the same manner as in Fig. \ref{fig:gain}.
    A: The schematic image of the strength change in the model.
    B: The overlaps of the fast dynamics with the target patterns are plotted for $\gamma=0.8,1,$ and $1.2$ in the left, center, and right panels, respectively.
    The color code for the patterns is the same as shown in D and E.
    C: The period of the sequence is plotted for varying $\gamma$.
    Circles and squares represent the two realizations of the networks that are the same shown in F.
    D: The dwell time for each pattern of the network shown in B is plotted for varying $\gamma$ in the top panel, as well as the transition time in the bottom panel.
    E: The stability for each pattern of the same network in shown B is plotted.
    F: The relation between the stability and period for the two networks is plotted. 
    }
    \label{fig:inp_str}
\end{figure*}

\subsection*{Input strength decreases the speed of the generation of sequence}
Next, we explore if the strength of the task input modulates the period of the sequential patterns (Fig. \ref{fig:inp_str}A).
To investigate this point, we change the strength of the context input, $\gamma$, after the learning with $\gamma_\text{lrn}=1$ is finished.
Figure \ref{fig:inp_str}B exhibits the neural dynamics with $\gamma$ varied in the same manner as in Fig. \ref{fig:gain}B.
For $\gamma=\gamma_{\text{lrn}}$, the five patterns are generated sequentially, as shown in Fig. \ref{fig:inp_str}B(ii).
In contrast to the gain modulation, the period of sequence is shorter for both smaller $\gamma (=0.8)$ and larger $\gamma=1.2$. 
As $\gamma$ is varied, the period takes a bell-shaped curve whose peak is located at $\gamma=\gamma_\text{lrn}$, as shown in Fig. \ref{fig:inp_str}C. 
Thus, the change in the input strength shortens the period, as compared with that for $\gamma=\gamma_\text{lrn}$.

We analyze the stability of the patterns as well as the dwell and transition time.
Here, the stability against the change in $\gamma$ is calculated in the same way as for the gain parameter.
The dwell and transition times for the five patterns are plotted in Fig.\ref{fig:inp_str} as functions of $\gamma$.
The dwell time curves peak around $\gamma=1$ ($=\gamma_{\text{lrn}}$) for all patterns.
The transient time is almost zero, except for the much higher $\gamma$. 
The anti-correlation between dwell and transient time is relatively unclear compared to the gain modulation, probably because the transient time is too small to see the clear correlation.
At the same time, as expected, the stability shows a similar curve with the dwell time curve, as shown in Fig. \ref{fig:inp_str}E.

The relationship between the stability and the period of patterns is plotted in Fig. \ref{fig:inp_str}F, showing that the period is highly correlated with the stability.
Further, we found that the dependence of the period on the stability differs below and beyond $\gamma=\gamma_\text{lrn}$;
for the same stability value, the period is higher for $\gamma>\gamma_\text{lrn}$ than for $\gamma<\gamma_\text{lrn}$, indicating that the period is controlled by the input strength differently for $\gamma>\gamma_\text{lrn}$ and $\gamma<\gamma_\text{lrn}$.

%Additionally, we analyzed the period modulation by top-down input in Supplemental Information.
%The results reveal that with the increase in the strength of the top-down input, the period is decreased, which is determined by the stability of the patterns.
All of these results show that the change in the period against the gain and  the task input is driven by the stability of the patterns in the sequence.
This mechanism coherently explains the speed modulations of sequential patterns by the different factors.

%%%%%%%%%%%%%%%%%%%%%%%%%%%%%%%%%%%%%%%%%%%%%%%%%%%%%%%%%%%%%%%%%%%%%%%%%%%%%%%%%%%%%%%%%%%
%%%%%%%%%%%%%%%%%%%%%%%%%%%%%%%%%%%%%%%%%%%%%%%%%%%%%%%%%%%%%%%%%%%%%%%%%%%%%%%%%%%%%%%%%%%

\subsection*{Transition time to the final state (reaction time) in the DMS tasks}
So far, we have shown that the flexible speed control of the sequence is achieved dominantly by the change in the dwell time through the stability modulation.
The transition time is simultaneously changed, but its change is much smaller than the dwell time and it is almost negligible for the dynamics speed control of the sequence  for the changes in the gain and input strength.
However, is the transition time not relevant to any speed control?
Actually,  it is experimentally observed that the transition time, rather than the dwell time, is relevant to cognition in the working memory task\cite{Ponce-Alvarez2012} and the perception task\cite{Benozzo2021}.
In these tasks, the sequential changes of metastable patterns are often observed and the transition time to a certain pattern, corresponding to the final decision state, is modified by expectation and the task difficulty, indicating the link between the cognition and the transition to the state.

\begin{figure}[h]
    %\centering
    \includegraphics[width=1\linewidth]{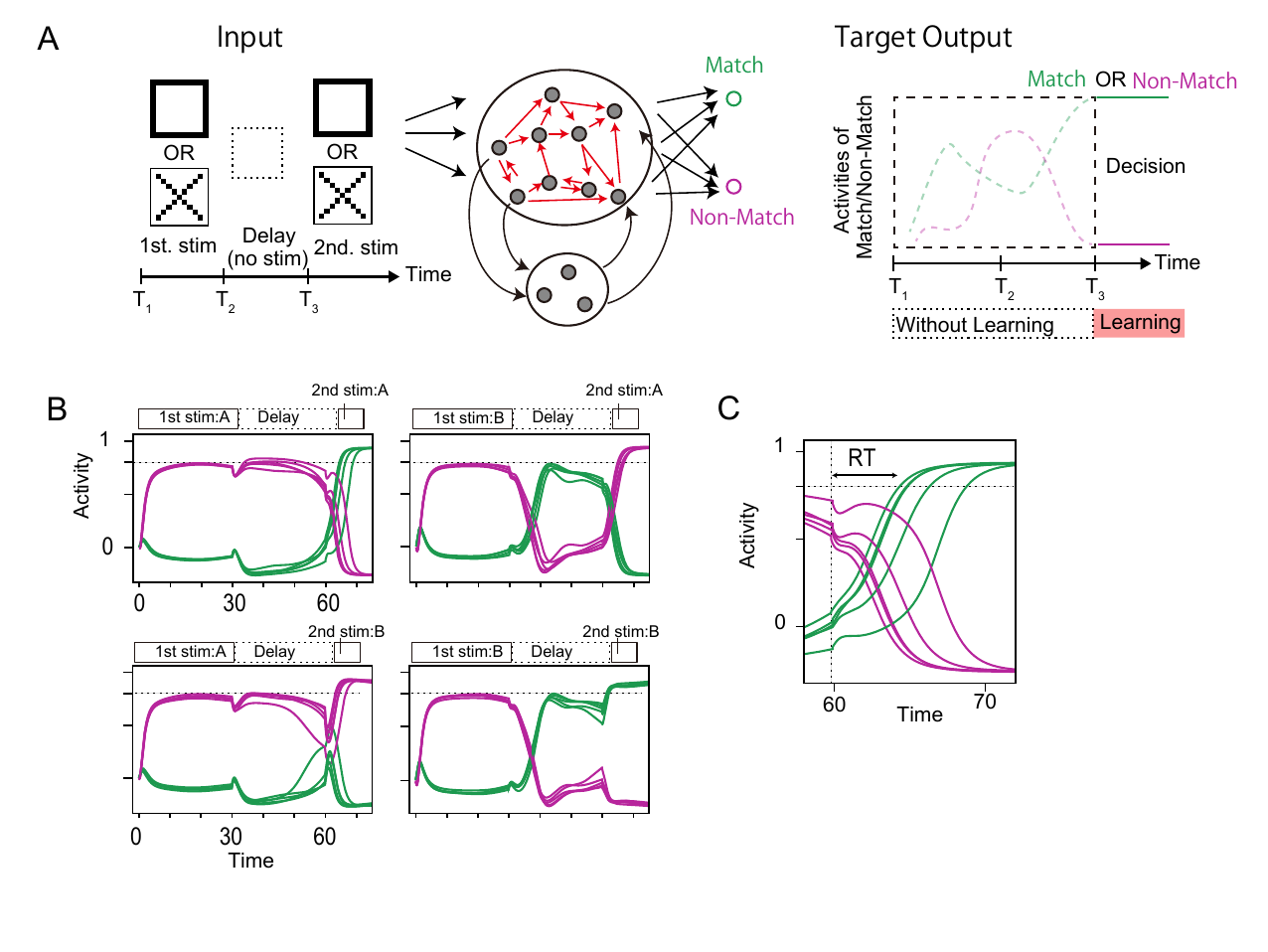}
    \caption{A Delayed Match-to-Sample (DMS) task in our model.
    A: Schematic image of the DMS task. There are two random patterns A and B for stimuli; A illustrated by a square image and B by a cross image for easy viewing. The first stimulus (randomly selected pattern out of A and B) is applied, followed by delay time.
    Then, the second stimulus (randomly selected again) is applied.
    The network is trained to generate the correct output pattern ("Match" or "Non-Match") based on these two successive input patterns (see Methods for details).
    The output is formally represented by the neural activities of "Match" and "Non-Match" neurons.
    The activities used for the training are those after the onset of the 2nd stimulus at $t=T_3$. 
    B: The output activities after learning are plotted for four conditions as examples.
    Five trials from different initial states are shown.
    The green and magenta color lines show the activities of Match and Non-Match neurons, respectively.
    C: The definition of the reaction time (RT) is illustrated.}
    \label{fig:DMS_image}
\end{figure}

\clearpage

Now we ask if such change of the transition time is related to the stability of the final state.
Here, we especially focus on the change in the working memory task, i.e., the delayed match to sample (DMS) task.
It is a typical working memory task,  where the two patterns are applied to subjects in succession with delay time.
The subjects are required to retain the first pattern and then to choose "match" or "non-match" after the second one is given when these two patterns are the same or not.

Before addressing the relation between the change in the transition time and the stability,  we first explain a neural network model for DMS task that is built by modifying our neural network model analyzed so far (See also the model details in Models).
Our model has the slow variables that can store the information of the first stimulus to choose the decision correctively.
In fact, this model have been demonstrated to perform DMS\cite{Kurikawa2021b}.
In contrast to our model, usual models for sequential patterns\cite{Amari1972,Kleinfeld1986,Seliger2003,Gros2007,Russo2012,Recanatesi2015,Murray2017,Mazzucato2019,Wyrick2021,Gillett2024} do not implement such a memory maintaining the previous inputs for the DMS task.

In this model, there are two input patterns, input A ( illustrated as "square") and B ("cross").
We applied either input twice in succession with the interval (1st and 2nd stimuli are randomly chosen each time), as shown in Fig. \ref{fig:DMS_image}A.
The network is required to output "Match" or "Non-Match" after the delay time, depending on whether the 1st and 2nd stimuli are the same or not.
We formally implement "Match" and "Non-Match" output neurons to visualize the behaviors of the network with the overlap of the fast neural activities with  "Match" and "Non-Match" patterns. 
Although the neural dynamics and the learning rule are identical to those used for the sequence generation, the synaptic connections are modified only after the delay time in the DMS model.
The network can be trained to generate correct choices depending on two successive inputs by retaining information of the first stimulus in the slow variables. 
Figure \ref{fig:DMS_image}B demonstrates that the network generates the correct choices after learning.
At the onset of the second stimulus, the activity of the correct output neuron is increased rapidly to converge to the decision state.

We explore the reaction time that is defined as the time in which the state converges to the final  decision state after the onset of the second stimulus, as illustrated in Fig. \ref{fig:DMS_image}C.
We investigate two types of speed control by measuring RT (see Methods); modulation by the gain parameter and by the difficulty in the task.
We do not measure the dwell time on the decision state because there is no state next to the decision.

\begin{figure}[h]
    %\centering
    \includegraphics[width=1\linewidth]{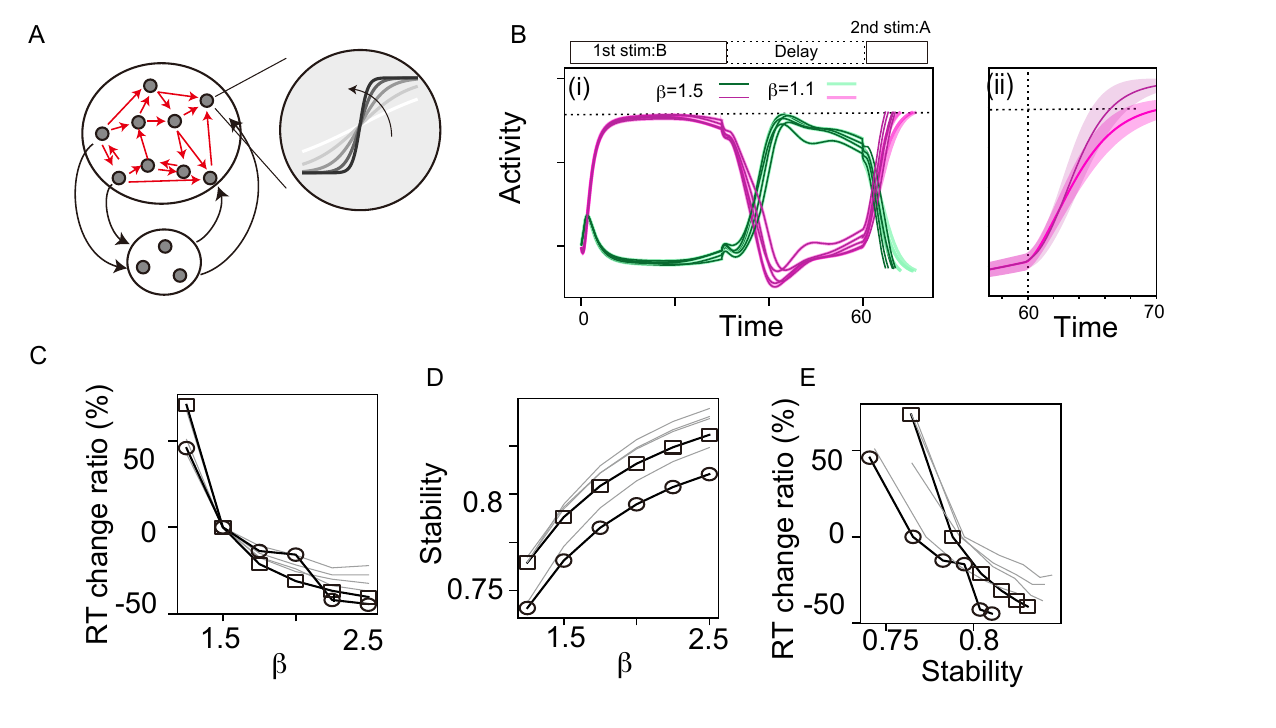}
    \caption{Change in the gain parameter $\beta$ in the DMS task.
    A: Schematic image of the change in $\beta$ in the model.
    B: The output neural dynamics for $\beta=1.1$ and $\beta=1.5$.
    The neural activities for different $\beta$ are superimposed for five trials in each value of $\beta$. 
    The panel (ii) is the expansion of  (i).
    The dotted lines indicate the threshold value for reaction time (RT) at $0.8$.
    C and D: RT change and the stability are plotted as functions of $\beta$ in C and D, respectively.
    Square and circles represent the two networks that are the same in C, D, and E.
    E: The relation between the stability and the RT change ratio is plotted.
    RT  and the stability are averaged values over 10 trials with different initial states. 
    Six networks with different learned patterns are shown in different lines in C, D, and E.
    }
    \label{fig:DMS_beta}
\end{figure}

\subsection*{Reaction (transition) time is changed by the gain parameter}
First, we analyze how the gain $\beta$ modulates the transition time to the final decision state, as shown in Fig. \ref{fig:DMS_beta}A.
Figure \ref{fig:DMS_beta}B demonstrates the dynamics of the match and non-match neurons for $\beta=1.1$ and $\beta=1.5$.
After the delay time, the activity of the non-match neuron (the correct output in this case) surges and reaches the decision threshold for both values of $\beta$.
On closer analysis, as shown in Fig. \ref{fig:DMS_beta}B(ii), the activity for $\beta=1.5$ reaches the threshold faster than that for $\beta=1.1$, suggesting that the reaction time is shorter for the larger $\beta$.

To investigate the dependence of the reaction time on $\beta$ comprehensively, we measured the reaction time for $\beta$ varied for different realizations of the networks and plot them in Fig. \ref{fig:DMS_beta}C.
As expected, the reaction time is reduced as $\beta$ increases.

The modulation of the reaction time suggests a change in the stability of the final decision state.
We examine whether the stability is related to the reaction time.
The stability of the final state is measured in the same way as that in the simple sequence generation and plotted in Fig. \ref{fig:DMS_beta}D.
We found that the stability is highly anti-correlated to the reaction time change, as shown in Fig. \ref{fig:DMS_beta}E.

\begin{figure}[h]
    %\centering
    \includegraphics[width=0.9\linewidth]{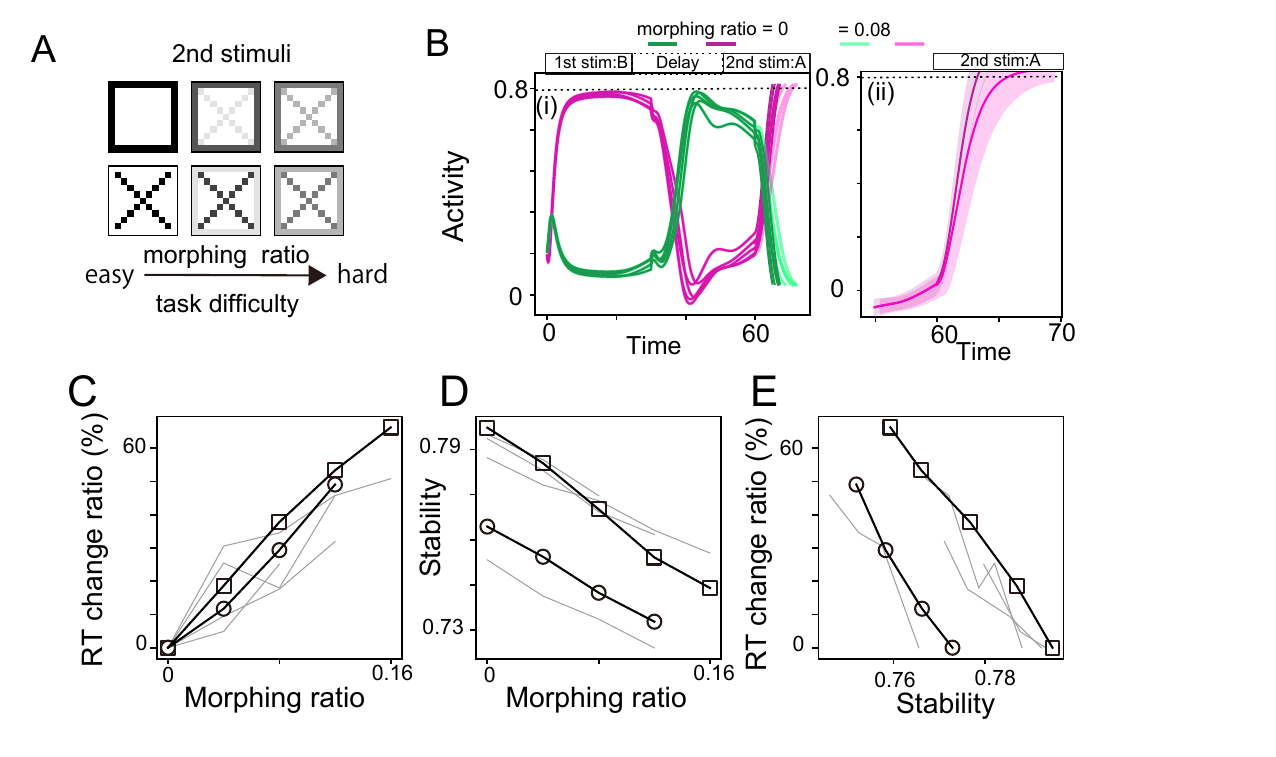}
    \caption{Change in the task difficulty shown in the same way as in Fig. \ref{fig:DMS_beta} with replacing $\beta$ by morphing ratio.
    A: Schematic image of change in the difficulty (morphing ratio).
    For changing the task difficulty, the applied patterns are modified:
    the similarity between the patterns is reduced with increasing a morphing ratio (see details in Methods).
    The morphing ratio between two patterns controls the difficulty (the high morphing ratio indicates the more difficult task).
    B: The output dynamics for morphing ratio $0$ and $0.08$.
    The neural activities for different morphing ratios are superimposed for five trials in each value of the ratio. 
    Panel (ii) is the expansion of (i).
    The dotted lines indicate the threshold value for reaction time (RT) at $0.8$.
    The shaded area represents SD.
    C and D: RT change and the stability are plotted against the change in the morphing ratio, respectively.
    E: The relationship between the stability and the change in RT is plotted.
    Square and circles represent the two networks that are the same in C, D, and E.
    RT  and the stability are averaged values over 10 trials with different initial states. 
    Six networks with different learned patterns are shown in different lines in C, D, and E.}
    \label{fig:DMS_diff}
\end{figure}

\subsection*{Reaction (transition) time is changed by the task difficulty}
Next, we examine whether task difficulty alters the transition time to the decision state.
Here, we implement task difficulty in the DMS task by using ambiguous stimuli (see Models for details):
An ambiguous stimulus is generated by morphing one of two learned patterns into the other as a second stimulus in test trials.
If the second stimulus is more ambiguous, the morphing ratio is larger and the task is more difficult, as shown in Fig. \ref{fig:DMS_diff}A.

The task difficulty changes the neural dynamics, leading to a longer reaction time.
The activity of the non-match neuron ramps up after the delay time, as demonstrated in Fig. \ref{fig:DMS_diff}B.
For high task difficulty,  the network requires more time to reach the threshold.
Actually, as the morphing ratio (namely, the task difficulty) increases, the average reaction time consistently increases despite the large fluctuations.
Such a behavioral trend is robust across different network realizations, as shown in Fig. \ref{fig:DMS_diff}
C.

Next, we ask if the task difficulty affects the stability of the decision state.
To answer this question, we measured the stability of the decision state by varying the morphing ratio and plot them in Fig. \ref{fig:DMS_diff}D, demonstrating that the stability decreases monotonically as the morphing ratio increases. 
In higher-difficulty tasks, the decision state becomes less stable because the applied 2nd input is more different from the optimal pattern used in learning.

We finally show that the reaction time is modulated by changing the morph rate through the stability, by measuring the relation between the reaction time and the stability.
Figure \ref{fig:DMS_diff}E displays the reaction time as a function of the stability, showing that the reaction time is anti-correlated with the stability.
All results reveal that increasing task difficulty prolongs the reaction time by reducing the stability of the decision state.

\begin{figure}[h]
    %\centering
    \includegraphics[width=1\linewidth]{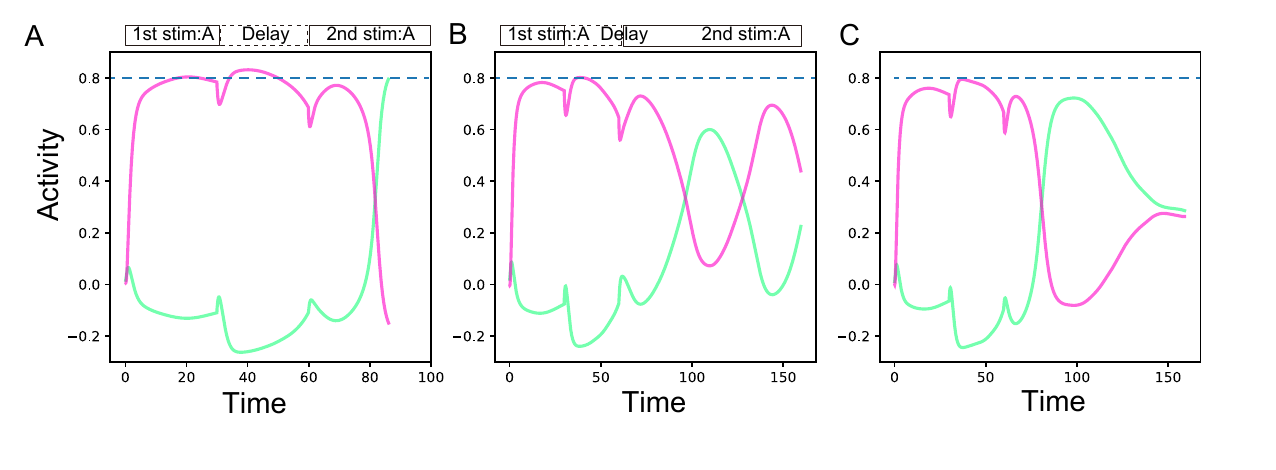}
    \caption{Typical behaviors of the neural dynamics showing wandering among the decision states.
    When the 2nd stimulus is applied, the network shows wandering behavior among the Match and Non-Match states before converging to the final state.
    The output activities of one sample trial are plotted for morphing ratio $0.12, 0.2$ and $0.24$ in A, B, and C, respectively.}
    \label{fig:DMS_wand}
\end{figure}

It is worth noting that in some trials for difficult stimuli, we observe that the neural activity wanders among the two choices, as exhibited in Fig. \ref{fig:DMS_wand}.
In this case, before the final decision (here, match choice),   the neural activity approaches the other choice (namely, non-match choice).
Such wandering among choices in neural dynamics is observed in  monkeys\cite{Rich2016} as subjective mind wandering, which is discussed later in Discussion.

\section*{Discussion}

%Sequence generation modelsの概要

%こまかいEI balanceでmetastaのstabilityが決まる(Vincent Fontanier,elife2022)
This study demonstrates that a variety of cognitive temporal modulations—changes in dwell time within a neural pattern or in reaction time to a decision state—can be explained by a unified principle: the modulation of stability in metastable neural states.
By using the recurrent neural network model with multiple timescales of neural activities, we investigated how the gain parameter in the activation function, input strength, and the task difficulty affect the speed of the sequential patterns in simple sequence and delayed match to sample task.
Specifically, we find that the gain enhances the stability of the patterns in the sequence and prolongs the dwell time, whereas the task difficulty decreases the stability and the transition (reaction) time to decision states.
Against the change in the input strength, the stability and the dwell time do not show the monotonic trend; 
If the input strength is smaller than that used in learning, it enhances the stability and prolongs the dwell time, and, oppositely, if it is larger, it declines the stability and decreases the dwell time.
In all cases, positive correlation between the stability and dwell time and the negative correlation between the stability and the transient time are consistently observed.

When performing cognitive tasks such as speech recognition, motor output, and context-dependent decision-making, task-specific sequential patterns are frequently observed in various cortical areas \cite{Jones2007,Miller2010,Shenoy2011,Mante2013,Kurikawa2018}. 
These patterns are influenced by numerous factors in diverse ways.
Top-down modulation, such as attention and expectation, alters the temporal characteristics of these sequential patterns\cite{Engel2016,Mazzucato2019,Wyrick2021}. 
Specifically, attention prolongs the duration of individual patterns within a sequence \cite{Engel2016}, while expectation accelerates convergence to stimulus-related patterns following external stimuli\cite{Wyrick2021}. 
Additionally, task difficulty has been shown to affect these temporal dynamics. 
For instance, when conditions in discrimination tasks are more challenging, the transition to decision-related patterns is delayed compared to easier conditions \cite{Ponce-Alvarez2012}. 
Similarly, the duration of sequential patterns is extended in incorrect trials compared to correct ones \cite{Benozzo2021}. 
Furthermore, contextual information, such as the length of delay intervals in working memory tasks\cite{Hardy2018,Wang2018,Zhou2023}, also modulates the speed of these sequential patterns while keeping the shape of the trajectories of neural dynamics.

Our study proposes that these diverse temporal modulations arise from variations in the stability of metastable patterns within sequences. 
In our computational model, gain modulation—which was commonly associated with top-down modulation in previous studies\cite{Mazzucato2019,Wyrick2021}—regulates the stability of these patterns, thus influencing both the dwell time within sequential patterns and the transition time to decision states. 
Additionally, external inputs, often represented as the context cues in recurrent neural network models \cite{Hardy2018,Wang2018}, enhance stability and consequently alter the sequential pattern speed. 
In difficult conditions in the working memory task, our model predicts longer transition times to decision states.
Therefore, our findings present a parsimonious mechanism by which the brain flexibly adjusts temporal characteristics in sequential patterns.

Although numerous models have previously been proposed for generating sequential patterns\cite{Amari1972,Kleinfeld1986,Seliger2003,Gros2007,Russo2012,Recanatesi2015,Murray2017,Mazzucato2019,Wyrick2021,Gillett2024}, these models typically have limitations when attempting to demonstrate such a parsimonious mechanism. 
In most of these earlier models, the temporal properties of sequences are inherently determined but not dependent on the external input and previous experience. 
For example, in models employing temporally asymmetric Hebbian-type connections—where connections are determined by the correlation between successive patterns— dwell times and transition intervals are solely determined by their connectivities. 
Similarly, models with clustered networks involve stochastic transitions, thus making the dwell times probabilistically pre-determined. While some studies have individually examined the effects of altering input strength\cite{Gillett2024,Murray2017} or modifying gain parameters \cite{Mazzucato2019,Wyrick2021} within these models, they remain limited in addressing sequential pattern modulations arising from history-dependent cognitive tasks, such as working memory, or discrimination tasks.

Beyond models with given connectivity, some studies employ recurrent neural networks trained via machine learning techniques \cite{Hardy2018,Wang2018,Zhou2023}. Although these networks can perform a variety of cognitive tasks effectively, they require global information during training—a condition biologically implausible in neural systems. Furthermore, these machine learning-based models fail to capture speed modulation of sequential patterns unless they are explicitly trained on both fast and slow sequences, which contradicts empirical observations.
Our model uses the simple Hebb-type learning, in which only pre- and post-synaptic information is required, and performs the speed modulation by learning only one condition ($\gamma_{lrn}=1$).
Thus, it overcomes these limitations and demonstrates a parsimonious and biologically plausible principle that allows for the flexible modulation of temporal dynamics across various cognitive tasks.

In addition to the modulation of the dwell and transition time, we found that the ambiguous stimulus (stimulus difficult to discriminate) leads to the neural state wandering among patterns corresponding to possible decisions before the final decision in the DMS task.
Interestingly, such wandering between states corresponding to possible choices has been reported in the orbitofrontal cortex in monkeys in the value-based alternative choice task\cite{Rich2016}.
Particularly, the wandering has been selectively observed in deliberation trials, 
suggesting that the monkeys might compare the values of choices.
Our finding suggests that, in deliberation trials, the stability of the decision state is somehow decreased, leading to the wandering neural dynamics.

Some previous modeling studies\cite{Mazzucato2019,Wyrick2021} have reported that the smaller gain in activation functions results in faster "state changes."
In contrast, our findings indicate that the higher gain leads to faster transitions to decision states. 
This apparent discrepancy may stem from the difference between the state change in previous studies and the transition in our study.
In the previous studies, the authors measured the time in which the neural state escaped from a metastable state to another state after the application of the input.
Thus, the measured time is a sum of the escape time and the convergence time to the final state.
The escape time is shorter and the convergence time is longer for the smaller gain in our model.
If the escape time is dominant, the total change time could be shorter for the smaller gain, consistent  with our results showing that it decreases 
In our model, in contrast, the measured transition time is purely convergence time to the final state after the onset of the second stimulus, because the neural state does not dwell on any stable state before the onset of the stimulus.
Therefore, the measured time in our study is shorter for the larger gain.
Which of the escape and the convergence times is dominant might depend on specific tasks.
Hence, the discrepancy may be resolved.

Finally, we discuss the functional role of the multiple timescales in neural activity in flexible temporal modulation.
A distribution of neural timescales has been extensively reported\cite{Golesorkhi2021,Cavanagh2020} across different cortical areas \cite{Hasson2015,Murray2014} and even within individual areas such as dorsolateal prefrontal cortex and orbitofrontal cortex\cite{Cavanagh2016,Cavanagh2018,Cavanagh2020,Wasmuht2018,Spitmaan2020}.
Although, as their functional role, Bayesian inference\cite{Ichikawa2024}, faster learning\cite{Perez-Nieves2021}, and representations of different task-relevant\cite{Kurikawa2024} are proposed,
our study suggest a novel functional role in temporal processing: these multiple timescales underpin for the flexible temporal modulation through the stability change.
Actually, fast neural dynamics give rise to metastable states, while slower dynamics serve to retain task-relevant information—such as the first stimulus in the delayed match-to-sample (DMS) task—and change the stability of the fast metastable states. 
The multiple timescales of neural activity allow us to coherently understand diverse temporal modulations not only in simple sequence generation, but also in complex cognitive tasks.

To summarize, our findings provide a unified and parsimonious explanation for the flexible modulation of the dwell and transient time observed in various cognitive tasks, as well as the wandering neural dynamics.
This contributes valuable insights into the neural mechanisms underpinning cognitive modulation.

% classical attractor model (or diffusion model)との対比は必要化か？
\begin{comment}
In cognitive science, In more challenging conditions, the neural system delays its response, possibly reflecting the need for additional integration of sensory evidence or increased competition between multiple decision states.  Attractor view. 
Our model provides the new insights that the ambiguous stimuli reduces the stability of the decision state.

The results suggest that decision state stability is not only a consequence of network dynamics but also a functional mechanism that adapts behavior to task demands. When the decision state is unstable, the network hesitates, reflected in slower reaction times. This adaptive mechanism might ensure that more difficult tasks are met with caution, improving the accuracy of responses at the cost of speed. Conversely, in easier tasks, the decision state stabilizes rapidly, enabling fast responses. Thus, task difficulty modulates reaction time by impacting the stability of decision states, with the fast and slow neural variabless dynamically balancing speed and accuracy based on task demands.
\end{comment}

%\section*{Acknowledgment}

\bibliography{15th}

\end{document}